# Electrically tuned magnetic order and magnetoresistance in a topological insulator

Zuocheng Zhang[1,*], Xiao Feng[1,2,*], Minghua Guo[1], Kang Li[2], Jinsong Zhang[1], Yunbo Ou[2], Yang Feng[1], Lili Wang[1,2], Xi Chen[1], Ke He[1,2,†], Xucun Ma[1,2], Qikun Xue[1], Yayu Wang[1,†]

[1]*State Key Laboratory of Low Dimensional Quantum Physics, Department of Physics, Tsinghua University, Beijing 100084, P. R. China*

[2]*Institute of Physics, Chinese Academy of Sciences, Beijing 100190, P. R. China*

\* *These authors contributed equally to this work.*

† Email: kehe@aphy.iphy.ac.cn; yayuwang@tsinghua.edu.cn

**The Dirac-like surface states of the topological insulators (TIs) are protected by time reversal symmetry (TRS) and exhibit a host of novel properties[1-3]. Introducing magnetism into TI, which breaks the TRS, is expected to create exotic topological magnetoelectric effects[4-8]. A particularly intriguing phenomenon in this case is the magnetic field dependence of electrical resistance, or magnetoresistance (MR). The intricate interplay between topological protection and broken-TRS may lead to highly unconventional MR behaviour that can find unique applications in magnetic sensing and data storage[9]. However, so far the MR of TI with spontaneously broken TRS is still poorly understood, mainly due to the lack of well-controlled experiments. In this work, we investigate the magneto transport properties of a ferromagnetic TI thin film fabricated into a field effect transistor device. We observe an unusually complex evolution of MR when the Fermi level ($E_F$) is tuned across the Dirac point (DP) by gate voltage. In particular, MR tends to be positive when $E_F$ lies close to the DP but becomes negative at higher energies. This trend is opposite to that expected from the Berry phase picture for localization, but is intimately correlated with the gate-tuned magnetic order. We show that the underlying physics is the competition between the topology-induced weak antilocalization and magnetism-induced negative MR. The simultaneous electrical control of magnetic order and magneto transport facilitates future TI-based spintronic devices.**

The studies of MR in novel magnetic materials have been at the center stage in the field of spintronics[9]. The most notable example is the discovery of giant magnetoresistance (GMR) when two ferromagnetic (FM) films separated by a nonmagnetic spacer layer have anti-parallel alignment[10]. GMR not only reveals the remarkable consequence of spin-dependent transport, but also caused a revolution in information technology. In a naïve picture, a TI film with long-range

FM order can be viewed as a bulk ferromagnet sandwiched between two layers of Dirac fermions with opposite chiralities, as illustrated in Fig. 1a. The MR behaviour in such a system is apparently of tremendous interest, but experimental challenges associated with magnetically doped TI have hindered a thorough exploration of this phenomenon. For example, recent studies show that Cr doped $Bi_2Se_3$, which is presumably the best TI, remains paramagnetic (PM) in spite of increased Cr content[11]. Transition metal doped $Sb_2Te_3$ exhibit robust FM order[12,13], but the charge transport is dominated by bulk carriers rather than the topological surface states. The most ideal situation, of course, is to have a TI film with *spontaneous* FM order and *tunable* $E_F$ over a wide doping range across the DP. In this work we achieve these goals by growing magnetically doped TI film with chemical formula $Cr_{0.15}(Bi_{0.1}Sb_{0.9})_{1.85}Te_3$ and fabricating it into a field effect transistor (FET) device, as schematically drawn in Fig. 1b. The mixing of Bi/Sb can tune the band structure close to an ideal TI[14], and when doped with Cr the system exhibits bulk FM order with out-of-plane easy axis[15]. The electrical gate can drive $E_F$ to various regimes of the TI band structure, leading to a rich evolution of MR that will be discussed below. Details about sample growth, device fabrication and measurements are presented in the Method session and Supplementary Information (SI).

We first demonstrate the FM order in the sample. Fig. 1d displays the temperature ($T$) dependent two-dimensional (2D) Hall resistivity ($\rho_{yx}$) measured at varied gate voltages ($V_g$). At the lowest $T$ = 1.5 K, the square-shaped hysteretic Hall traces for all $V_g$s indicate robust FM order. The total $\rho_{yx}$ in this case can be expressed as $\rho_{yx} = R_A M + R_H H$. Here $M$ is the magnetization of the sample and $R_A$/$R_H$ is the anomalous/ordinary Hall coefficient respectively. With increasing $V_g$ from -15 V to +15 V, the $R_H$ at $T$ = 1.5 K changes from positive to negative, indicating the change from *p*- to *n*-type charge carriers via the bulk insulating regime. The existence of robust ferromagnetism across the entire $V_g$ range is similar to that observed in Cr doped $(Bi,Sb)_2Te_3$ films with varied Bi/Sb ratio[15]. The FM order becomes weaker with increasing $T$ and disappears for all $V_g$s at $T = 15$ K, when the Hall curves become reversible but retain strong nonlinearity due to strong FM fluctuations.

The gate tuning of carrier density can be best demonstrated by the Hall measurements at $T = 80$ K (see SI for details), when the influence of FM fluctuation is much reduced. Fig. 1c shows the $V_g$-dependent nominal carrier density $n_H$ estimated from the ordinary Hall coefficient $R_H$ as $n_H = 1/R_H$. The most dramatic feature here is the divergent behaviour around $V_g = +2$ V, when the sample is in the charge neutral regime with $E_F$ lying close to the DP[16]. On the left (right) side, $E_F$ lies below (above) the DP with hole-type (electron-type) carriers tuned systematically by $V_g$. At the large $V_g$ limit close to $\pm 15$ V, the bulk band carriers start to dominate the charge transport. The same trend is revealed by the longitudinal resistivity $\rho_{xx}$ measured under varied gate voltages. As shown in Fig. 1e, at $V_g = \pm 15$ V the $\rho_{xx}$ shows a metallic behavior at high $T$ and becomes weakly insulating at low $T$. As $E_F$ is tuned towards the DP, the $\rho_{xx}$ value keeps rising and the high $T$ behavior gradually becomes insulating between $V_g = -2$ V and $+4$ V. These results are consistent with the Hall effect measurements and confirm that the bulk is insulating in this regime but becomes metallic at higher $V_g$.

We next turn to the MR behaviour of the FM TI film. Fig. 2a illustrates the $H$ dependent MR ratio, defined as $\dfrac{\delta\rho}{\rho_0} = \dfrac{\rho_{xx}(H) - \rho_{xx}(0)}{\rho_{xx}(0)}$, measured at $T = 1.5$ K under varied gate voltages (the raw $\rho_{xx}$ vs. $H$ curves are documented in Fig. S2 in SI). The butterfly-shaped hysteresis is characteristic of the negative MR in FM metals caused by the spin-dependent scattering of carriers by local magnetic ordering[10]. The peak position corresponds to the coercive field ($H_C$), and on either side of $H_C$ the reduced scattering of a specific spin orientation leads to negative MR. When $V_g$ is reduced to 0 from $\pm 15$ V, $H_C$ decreases monotonically. The MR behaviour at $T$ = 5 K (Fig. 2b) is similar, except that the $H_C$ value is much smaller (note the change of scale). Fig. 2c summarizes the $V_g$ and $T$ dependence of $H_C$ in a color contour plot. The general trend is a "V"-shaped pattern with apparent asymmetry, where the hole-doped side has stronger FM order. The Curie temperature $T_C$ can be estimated from the temperature when $H_C$ drops to 0, as shown by the thick blue line. The contour plot clearly demonstrates the enhanced FM order on both sides of the DP, when more carriers are injected into the film by electrical field. The same conclusion can be drawn from the contour plot of gate-tuned anomalous Hall effect (AHE) shown in Fig. S3 in SI. In fact this "V"-shaped gate voltage dependence of $H_C$ was already present, although totally overlooked, in our previous work on a similar magnetic TI[17]. As shown clearly in Fig. 2a of Ref. 17, $H_C$ is smallest around $V_g$ = -1.5 V, when $E_F$ lies in the magnetic gap at the DP and AHE is quantized. $H_C$ increases on either side of the DP (e.g. the $V_g$ = -55 V and 200 V curves apparently have larger $H_C$ than the $V_g$ = -1.5 V curve), in good agreement with the trend observed here.

The MR behaviour at higher temperatures is even more perplexing. Fig. 3a displays the $V_g$-dependent MR curves measured at $T$ = 8 K. The $V_g$ = -15 V curve still shows the negative MR similar to that observed at lower temperatures, except that the hysteresis is barely visible due to much reduced $H_C$. With decreasing negative $V_g$ to -2 V, the negative MR remains qualitatively the same although the amplitude keeps decreasing. At $V_g$ = 0 V, the MR is still negative at very weak $H$, but crossovers to positive for $H$ > 0.1 T. At positive $V_g$ up to +4 V, the MR exhibits a sharp positive increase with $H$, but at $V_g$ = +7 V the curve returns to the behaviour similar to that at $V_g = 0$ V with a negative dip at weak $H$ and a slow positive increase at higher $H$. With further increase of positive $V_g$ up to +15 V, the MR behaviour becomes the same to the regime with large negative $V_g$, showing strongly negative MR. This unusually complex evolution of MR with varied $V_g$ is also seen at higher temperatures. The main change with increasing $T$ is in general the positive MR becomes more pronounced.

Figure 4a and 4b summarize the $T$-dependent MR curves measured at two representative gate voltages. The position of $E_F$ corresponding to each $V_g$ is shown in the schematic band structure in Fig. 4c and 4d. It is clear that when $E_F$ lies way below the DP ($V_g = -15$ V) and cuts through the bulk valence band, the negative MR behaviour dominates at all temperatures. In contrast, when $E_F$ lies close to the DP ($V_g = +2$ V) and within the bulk band gap, MR becomes positive for the whole $H$ range for $T \geq T_C = 8$ K, as well as the large $H$ regime for $T < T_C$. Similar $T$-dependent MR curves measured at other $V_g$s are shown in Fig. S4. The overall evolution of MR can be directly visualized in Fig 4e, where a color scale is used to characterize the relative MR ratio measured at $\mu_0 H = 0.3$ T, which is beyond the $H_C$ of all curves to avoid complications from hysteresis. Interestingly, there is a close correlation between the MR ratio shown here and the $H_C$ plot in Fig. 2e, both showing a "V"-shaped pattern centered around $V_g = +2$ V. In particular, negative MR is favored in the regime with stronger ferromagnetism, both in terms of temperature and gate voltage. The same pattern can also be seen in the contour plot of MR taken at other magnetic fields (Fig. S5).

It is very puzzling why the MR has such a complex evolution with gate voltage. In conventional FM metals or diluted magnetic semiconductors (DMSs), the amplitude of MR may depend on carrier density[18] but a sign change is rare. Here we show that in a FM TI film the sign of MR can change *twice* as the gate voltage is varied between ±15 V at certain temperature range (Fig. 3a to 3c). Next we will unravel the mechanism of this peculiar behaviour, which represents a novel transport phenomenon in TIs with spontaneous FM order.

At first glance the results shown in Fig. 3 are reminiscent of the crossover between weak antilocalization (WAL) and weak localization (WL) proposed for magnetically doped TI[19], but a closer examination of the data suggests otherwise. In pristine TI, spin-momentum locking of the Dirac cone induces a $\pi$ Berry phase between two time-reversed self-crossing loops[20]. The destructive quantum interference between them leads to WAL, and magnetic field penetrating the loops suppresses the interference and causes a positive MR[11,21,22]. WAL indicates that the Dirac fermions cannot be localized by nonmagnetic impurities, which is a key aspect of the topological protection[1,3]. Magnetism introduced into TI breaks the TRS and opens a gap at the DP[23, 24]. The Berry phase $\phi$ in this case can be expressed as:

$$\phi = \pi(1 - \frac{\Delta}{2E_F}),$$

which depends on the gap size $\Delta$ and position of $E_F$ (Ref. [19]). When $E_F$ lies close to the DP, $\phi$ approaches zero, thereby MR should be negative due to WL. However, Fig. 3 shows that MR is positive for $E_F$ close to DP and becomes negative for $E_F$ at higher energies. This trend is totally opposite to that predicted by the Berry phase picture and indicates that novel physics beyond simple localization must be at work.

We gain key insight into this puzzle from the intimate relationship between MR and ferromagnetism. Let's first focus on the limit of $V_g = -15$ V, which has the strongest FM order. Here $E_F$ cuts through the bulk valence band so that bulk holes make dominant contribution to charge transport (Fig. 4c). At low temperatures ($T \leq 5$ K) the MR is negative with pronounced butterfly-shaped hysteresis, which is the same as that commonly observed in FM metals due to the spin-dependent scattering of charge carriers. With increasing temperature to above $T_C$, the long-range FM order vanishes but strong FM fluctuations make the physics still valid. As shown in Fig. 4a, from 1.5 K to 15 K the MR curves remain qualitatively the same except for the gradual disappearance of $H_C$. The negative MR of the bulk carriers in FM TI thus has little difference from the magnetism-induced negative MR in conventional FM metals, which is not totally unexpected. The irrelevance of the WL picture in this regime is further supported by the unphysical fitting parameters when we use the Hikami-Larkin-Nagaoka (HLN) formula for localization[23] to simulate the magnetoconductivity (MC) curves (see SI F for details).

As $E_F$ is moved towards DP by the gate voltage ($V_g = +2$ V), the charge transport predominantly comes from the surface states because $E_F$ lies in the bulk gap (Fig. 4d). At $T \geq T_C = 8$ K, there is no long-range FM order thus the Dirac cone is gapless. The positive MR in this regime is characteristic of the WAL of surface Dirac fermions, just like that in pristine TI. In this regime the MC curves can be fitted very well by the HLN formula with reasonable parameters showing the WAL behaviour (Fig. S6). Upon cooling to below $T_C$, however, the system shows the butterfly-shaped negative MR at weak $H$. Therefore even the surface Dirac fermions exhibit rather conventional spin-dependent scattering with local magnetization. As has been shown by recent angle-resolved photoemission spectroscopy measurements, the long-range FM order opens a gap at the Dirac point[24] and leads to a hedgehog spin texture for states outside the magnetic gap[25]. The spins of the Dirac fermions in this regime obtain an out-of-plane component, and will suffer spin-dependent scattering with the local magnetization. Now the massive Dirac fermions behave more like conventional fermions due to the suppression of topological protection by broken TRS. Only at $H$ higher than $H_C$ does the MR recover the positive tendency, manifesting the WAL due to residual topological protection.

Therefore, the perplexing MR behaviour actually reflects the competition between topology-induced WAL and magnetism-induced negative MR. In the large $V_g$ regime where FM order is stronger and bulk conduction prevails, MR is negative in a similar manner to conventional FM metal. In the $V_g$ range when $E_F$ lies in the bulk gap and surface Dirac fermions are mainly responsible for charge transport, MR is positive due to the WAL effect except for the weak field regime in the FM state. The MR behaviour at intermediate $V_g$ can be described as the competition of these two effects (Fig. S3). The versatile electrical control of ferromagnetism and MR also allow us to construct novel spintronic and topological magnetoelectric devices using magnetic TIs. For example, the fact that the sign of MR switches *twice* when $V_g$ is changed between $\pm 15$ V may be used to achieve novel magnetic readout and signal transmission functions. In certain sense this is analogous to the famous single electron transistor, which turns on and off again every time one electron is added to the quantum dot by $V_g$ (ref. 26).

The only remaining question is why the FM order is strengthened by increased carrier density (Fig. 2c). The origin of magnetism in TI has attracted much recent attention. Available proposals include the bulk van Vleck mechanism[27] and surface Dirac fermion mediated Ruderman-Kittel-Kasuya-Yosida (RKKY) mechanism[28]. The former one is insensitive to the carrier density[15] whereas in the latter one FM order becomes weaker when $E_F$ moves away from the DP[29]. Our gate-tuning results provide a new trend that has never been reported before. The enhanced FM order at heavily electron and hole doped regime, when the charge transport is dominated by bulk carriers, suggests that the RKKY mechanism involving itinerant bulk carriers is also essential. In fact it has been shown that in Cr doped $Sb_2Te_3$ the bulk valence-band holes plays an important role in ferromagnetism[13], but it has yet to be revealed if the bulk conduction-band electrons can do the same thing. Our experimental data shown in Fig. 2c suggest so, although there is an apparent electron-hole asymmetry. This is different from the well-known III-V group DMSs, in which only hole-type RKKY is available[30]. A plausible reason is that the TIs have inverted bulk band structure, in which the original bulk conduction/valence bands are hybridized thus share common characteristics. Therefore bulk electrons can mediate FM order in a similar manner to the bulk band holes, which is unique to the TIs. We note that further theoretical studies with realistic band structure of magnetically doped TI will be needed to confirm this hypothesis.

## Method Summary:

The $Cr_{0.15}(Bi_{0.1}Sb_{0.9})_{1.85}Te_3$ TI thin film with thickness of 5 quintuple layers (QL) is grown on insulating $SrTiO_3(111)$ substrate by using molecular beam epitaxy (MBE). The film is covered with 10 nm Te capping layer before being taken out of the MBE chamber. A 50 nm-thick amorphous $Al_2O_3$ film deposited by atomic layer deposition is used as the dielectric layer for the top gate. The sample is fabricated into a Hall-bar geometry by using photolithography. The length between the longitudinal contacts is 50 μm and width of the sample is 50 μm. Electric contacts are made by mechanically pressing Indium into the film. The 2D Hall resistivity $\rho_{yx}$ and longitudinal resistivity $\rho_{xx}$ are measured by using standard four-probe ac lock-in method with the magnetic field applied perpendicular to the film plane.

## Acknowledgements

This work was supported by the National Natural Science Foundation of China, the Ministry of Science and Technology of China, and the Chinese Academy of Sciences.

## Author Contributions

K.H., X.-C.M., Q.-K.X. and Y.W. designed the experiments. X.F., K.L. and Y.O. carried out MBE sample growth and device fabrication. Z.-C.Z, M.-H.G., J.-S.Z., and Y.F. carried out transport measurements. L.-L.W., X.C. and X.-C.M. assisted in the experiments. Z.-C.Z, K.H., Q.-K.X. and Y. W. prepared the manuscript. All authors have read and approved the final version of the manuscript.

## Additional information:

The authors declare no competing financial interests.

Figure Captions:

**Figure 1 | Measurement setup and electrical characterization of the field effect transistor.** a, A FM TI thin film consists of two layers of Dirac fermions separated by the bulk FM layer. The Dirac fermions residing on the two surfaces have opposite spin chirality. A gap is opened at the DP due to the out-of-plane ferromagnetism. b, Schematic device for the magnetotransport measurement of the $Cr_{0.15}(Bi_{0.1}Sb_{0.9})_{1.85}Te_3$ thin film. A Hall bar geometry is used to measure the $\rho_{xx}$ and $\rho_{yx}$ simultaneously under various gate voltages and magnetic field. c, Electric field dependence of carrier density at 80 K. Carrier density is estimated from the Hall curve at 80 K. The carrier density shows divergent behaviour around $V_g = +2$ V. d, Temperature dependence of the anomalous Hall effect under various gate voltages. Strong FM order is confirmed by the square-shaped hysteresis at 1.5 K over all the gate voltages. The ordinary Hall slope evolves from positive to the negative when the $E_F$ is tuned across the DP. e, Temperature dependence of 2D longitudinal resistivity under varied gate voltages. The high $T$ metallic behavior at large $|V_g|$ gradually becomes insulating as $E_F$ is tuned into the bulk gap by the gate voltage.

**Figure 2 | Magnetotransport results at low temperature and evolution of $H_C$ and $T_C$.** a,b, MR measured at $T = 1.5$ K (a) and 5 K (b). Each curve is offset by 3 percent (a) and 1 percent (b) for clarity. Butterfly-shaped hysteresis confirms the FM order at low temperature. The negative MR is due to local magnetic order induced spin dependent scattering. $T_C$ reaches a minimum at +2 V and increases on both sides, indicating enhanced FM order in both $p$- and $n$-type regime. The increasing temperature reduces the $H_C$ and relative change of MR. c, Summary of the $H_C$ and $T_C$ at various gate voltages. $T_C$ is defined as the temperature when $H_C$ reaches zero within

experimental uncertainty. The asymmetry behaviour of the $V_g$-dependent $T_C$ suggests that hole doped regime has stronger FM order.

**Figure 3 | Evolution between magnetism-induced negative MR and topology-induced WAL.** Gate voltage dependence of MR at various temperatures. At $T = 8$ K (a), magnetism-induced negative MR at -15 V gradually evolves to the topology-induced WAL behaviour as the gate voltage is swept to +2 V, but with further increase of $V_g$ up to +15 V the MR becomes negative again as that in the $p$-type regime. Similar trend is observed at $T = 10$ K (b), 12 K (c), and 15 K (d).

**Figure 4 | Correlation between negative MR and ferromagnetism.** a,b, Temperature dependence of MR at two representative gate voltages. (a) $V_g = -15$ V. As the $E_F$ cuts through the bulk valence band, negative MR is observed over all temperature. Each curve is offset by 1.5 percent for clarity. (b) $V_g = 2$ V. When the $E_F$ is tuned into the bulk gap, WAL begin to take charge the transport and WAL is observed above $T_C$. Each curve is offset by 1 percent for clarity. c,d, The position of $E_F$ at the two gate voltages $V_g = -15$ V (c) and $V_g = +2$ V (d). e, Color plot of the relative change of MR at $\mu_0 H = 0.3$ T. Negative MR is more pronounced in the regime of high $V_g$ and low temperature, which are favorable for the enhancement of ferromagnetism.

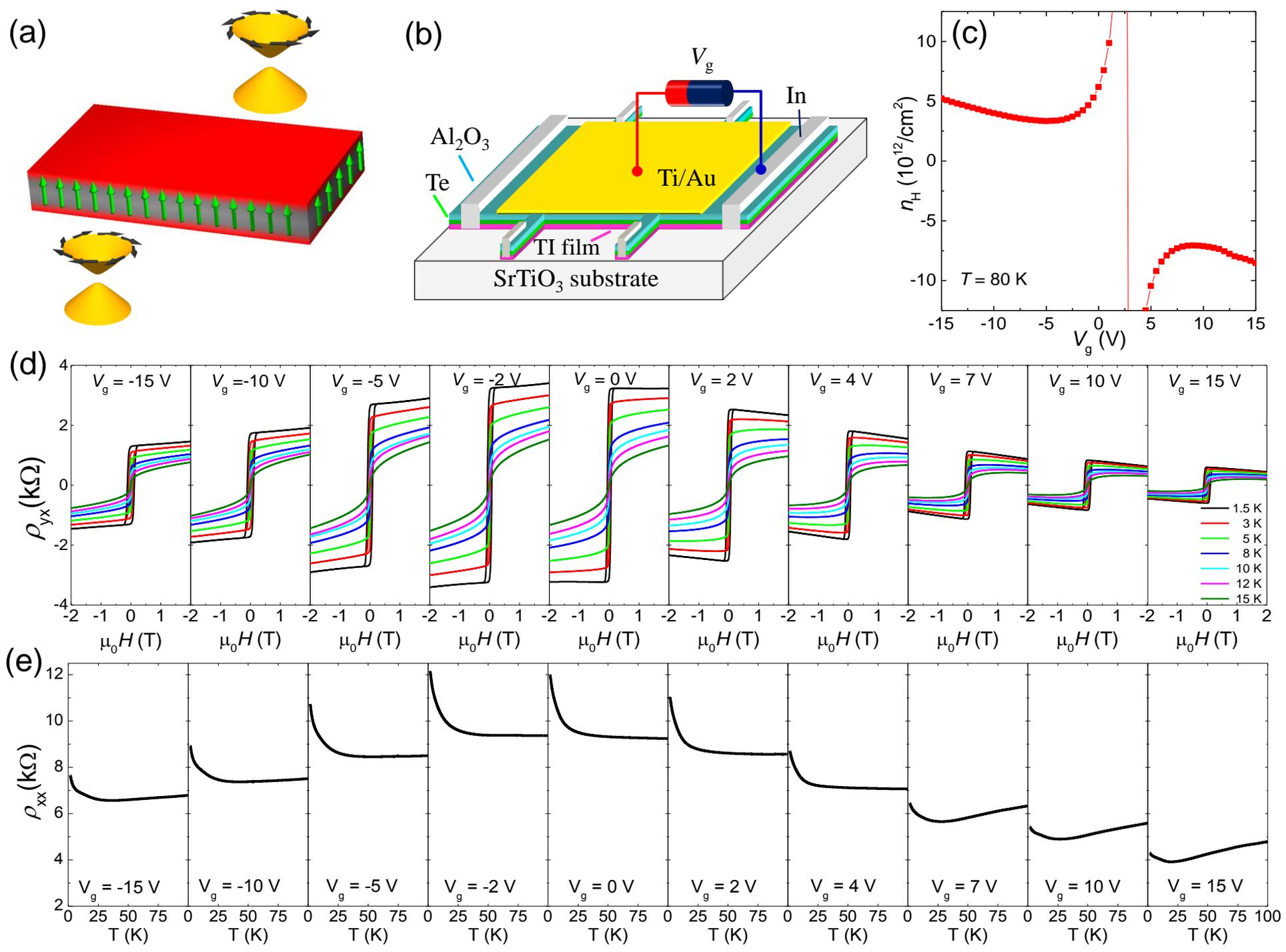

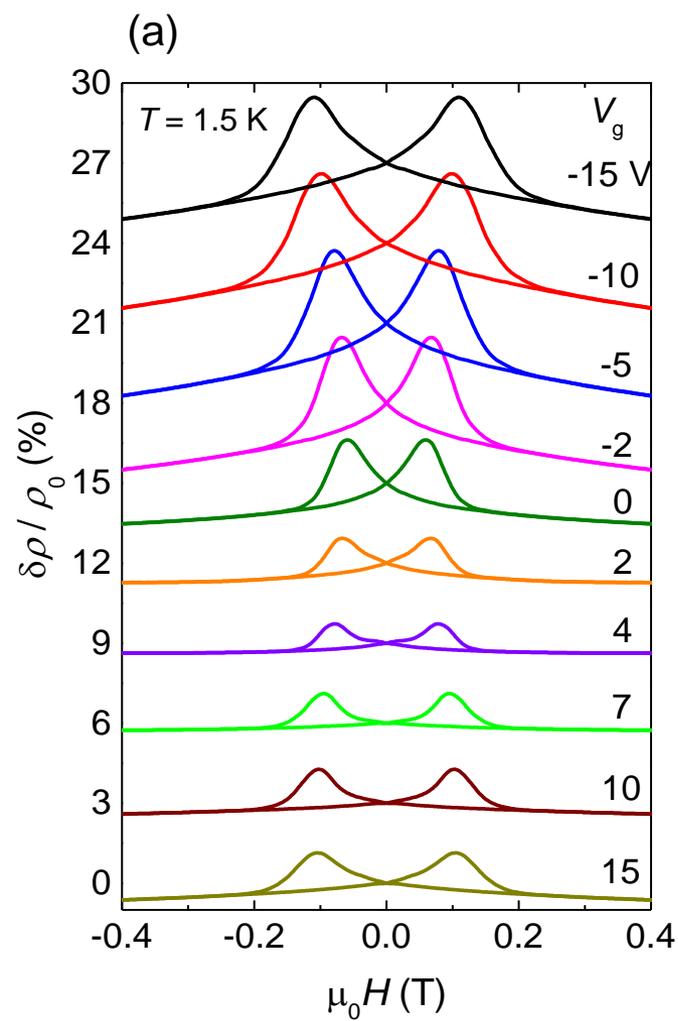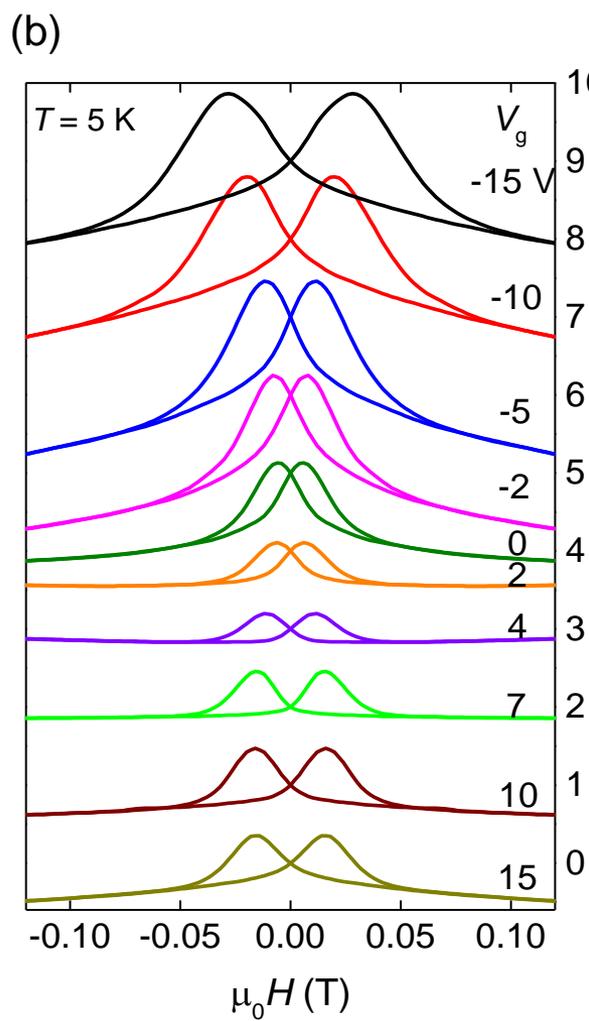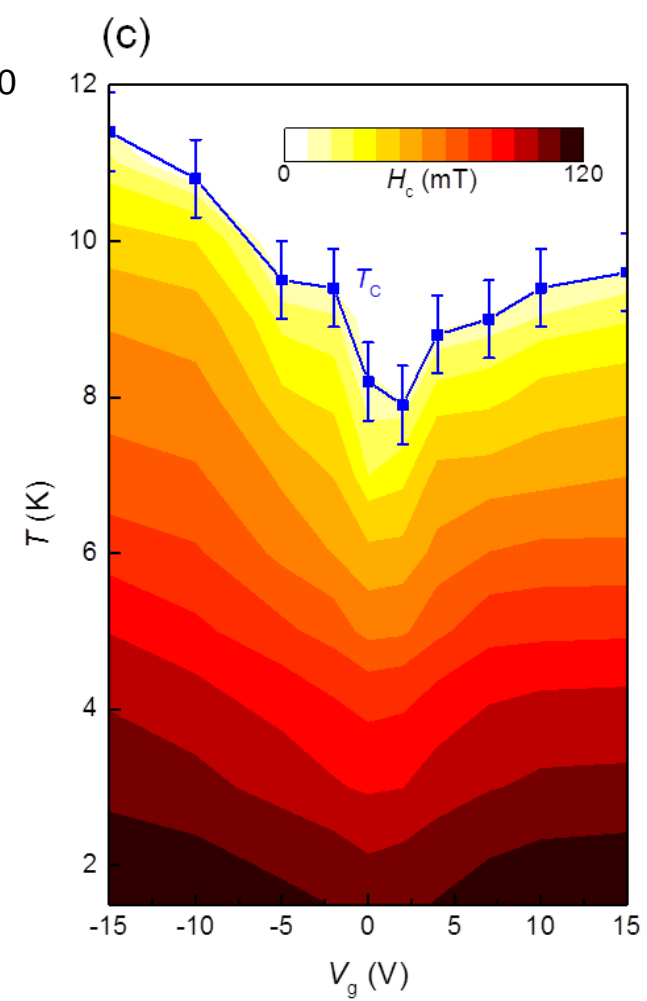

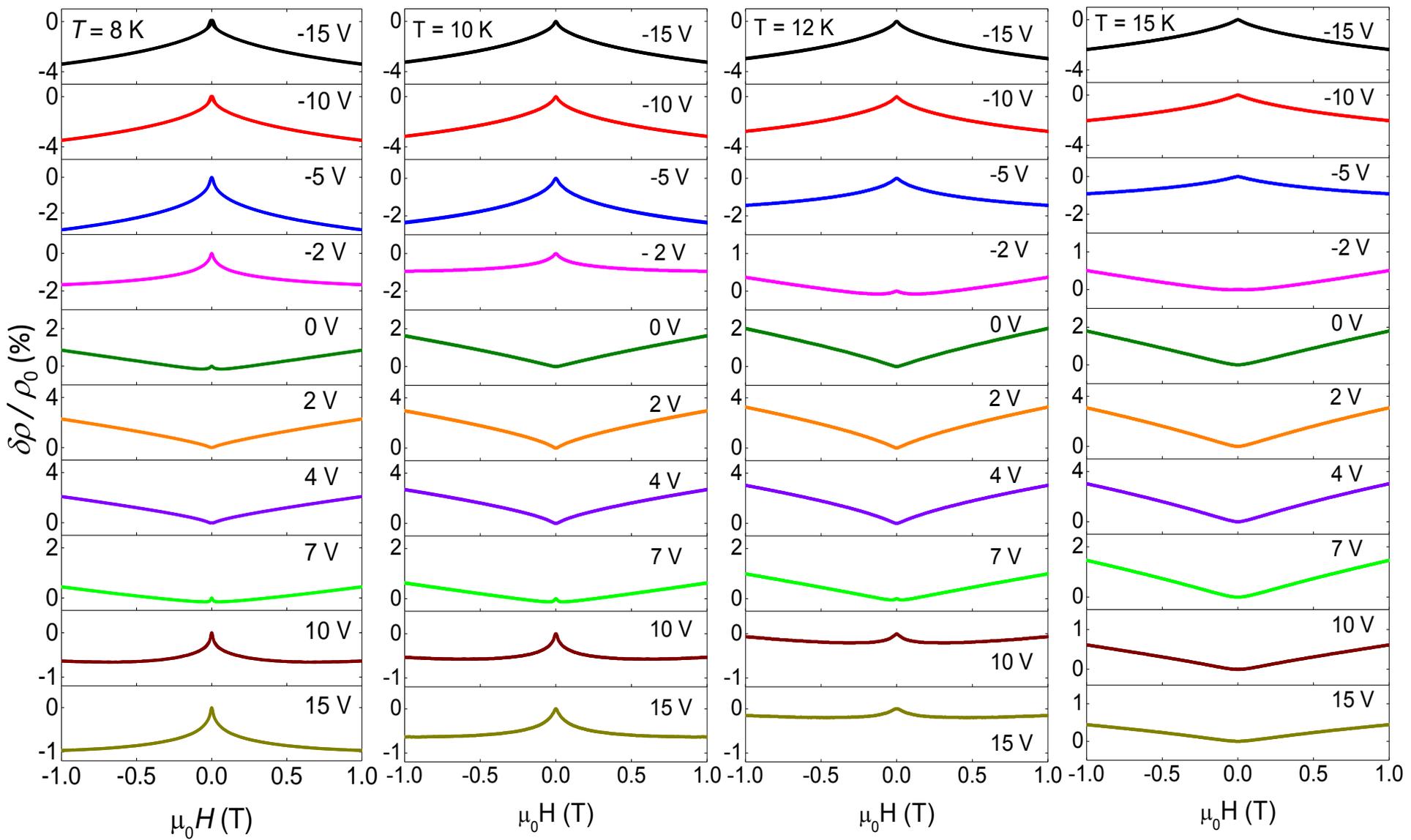

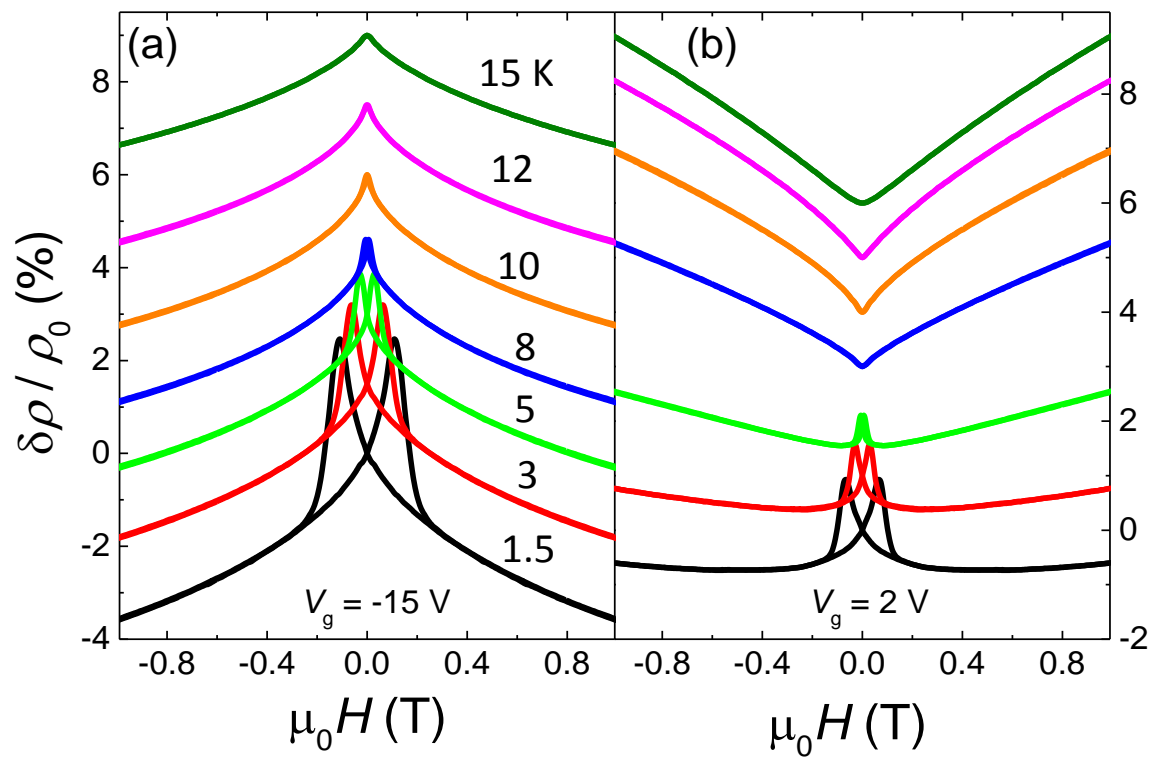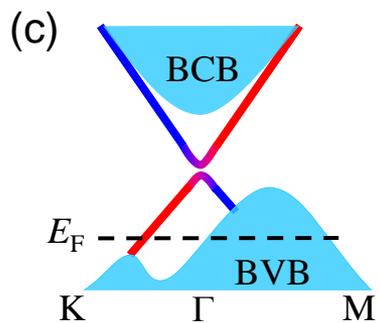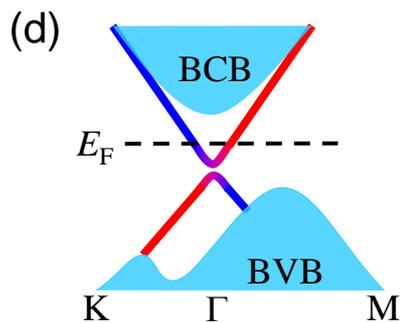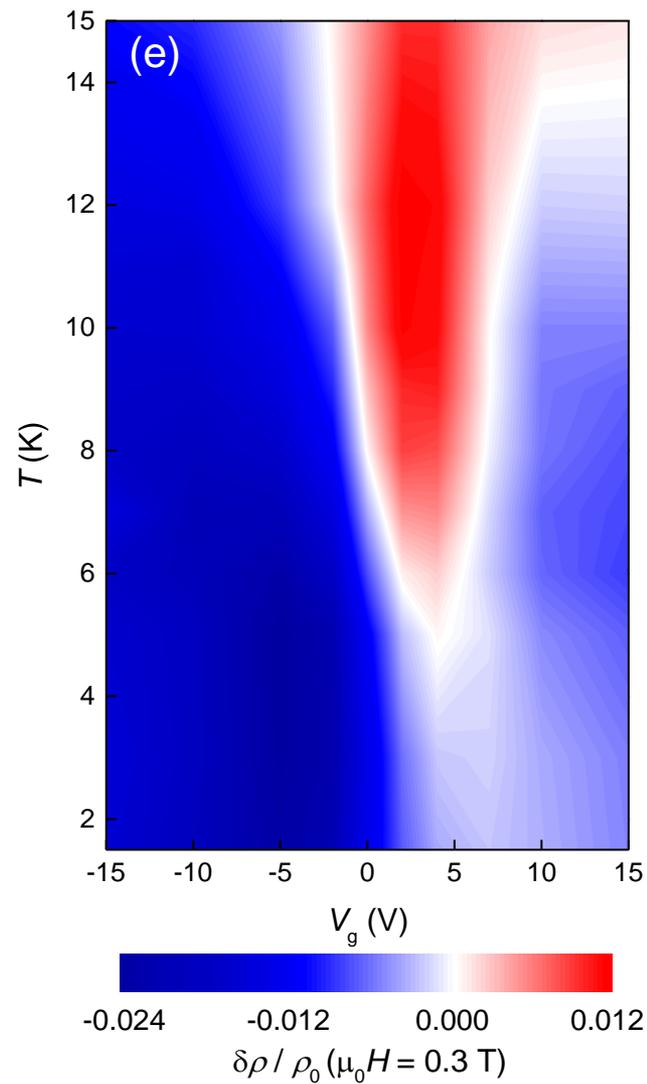

# Supplementary Information:

# Electrically tuned magnetic order and magnetoresistance in a topological insulator

Zuocheng Zhang[1,*], Xiao Feng[1,2,*], Minghua Guo[1], Kang Li[2], Jinsong Zhang[1], Yunbo Ou[2], Yang Feng[1], Lili Wang[1,2], Xi Chen[1], Ke He[1,2,†], Xucun Ma[1,2], Qikun Xue[1], Yayu Wang[1,†]

[1]*State Key Laboratory of Low Dimensional Quantum Physics, Department of Physics, Tsinghua University, Beijing 100084, P. R. China*

[2]*Institute of Physics, Chinese Academy of Sciences, Beijing 100190, P. R. China*

\* *These authors contributed equally to this work.*

† Email: kehe@aphy.iphy.ac.cn; yayuwang@tsinghua.edu.cn

**Contents:**

## A. Estimation of the carrier density

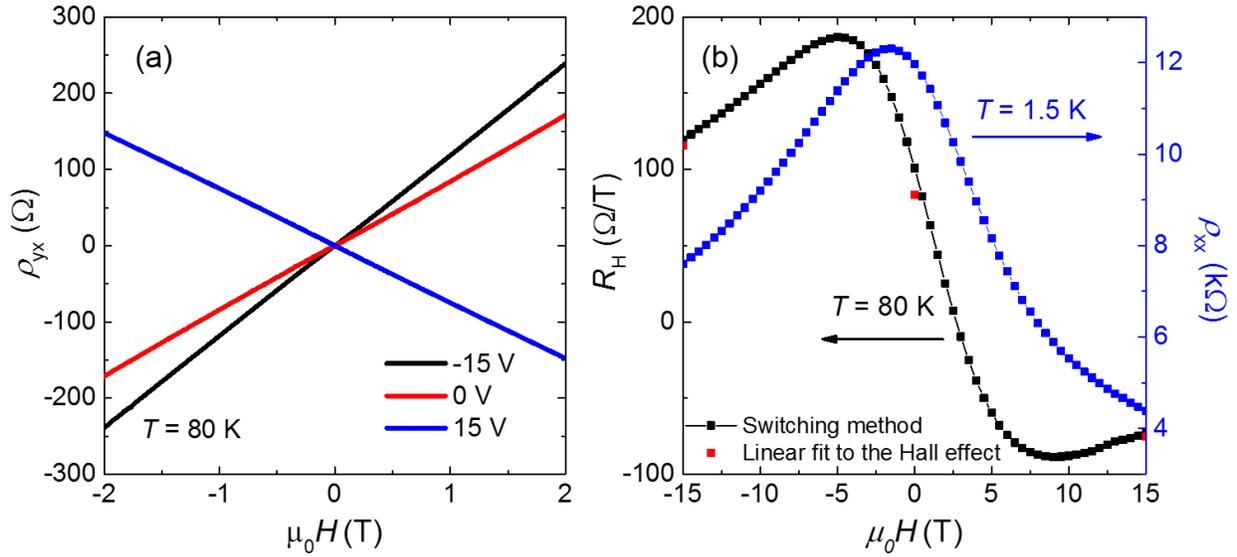

**Figure S1: Measurement of the ordinary Hall coefficient.** (a) The Hall resistivity as a function of magnetic field measured at $T = 80$ K under three representative $V_g$ = -15 V, 0 V and +15 V. All the Hall traces are straight lines, indicating the sole contribution of the ordinary Hall effect of charge carriers. (b) Electric field dependence of Hall slope at 80 K and zero magnetic field 2D resistivity at 1.5 K. The black symbols indicate the ordinary Hall coefficient measured by the switching method at fixed magnetic field of 2 T, which renders high density data point. It agrees well with the $R_H$ estimated from the slope of the straight lines in Fig. S1a (red squares). $\rho_{xx}$ shows a peak in the charge neutral regime.

At temperatures near the Curie temperature ($T_C$), ferromagnetic fluctuations contribute to the Hall effect signal and lead to strong nonlinearity to magnetic field ($H$). In order to have a reliable estimate of the carrier density, we measured the ordinary Hall effect at 80 K, which is much higher than the $T_C$ of the sample. As shown in Fig. S1a, at $T = 80$ K the Hall resistance is linear to $H$ at three representative gate voltages ($V_g$s). The Hall coefficient $R_H$ can be obtained from the slope of the straight lines, and the nominal carrier density can be calculated as $n_H = 1/R_H$. To get a high-density trace of $R_H$ vs. $V_g$, we use another method invented by Sample and coworkers [S1] to measure $R_H$. In this measurement the $\mu_0 H$ is fixed at 2 T, while the Hall effect is measured by switching between the voltage and current leads to extract the antisymmetric component of the resistivity tensor. The $V_g$ dependent Hall coefficient $R_H$ obtained by this way is plotted in Fig. S1b, and the nominal carrier density estimated from $R_H$ is plotted in Fig. 1c in the main text. The $R_H$ measured by this switching method is in good agreement with that obtained from the linear fit of the straight lines in Fig. S1a (the three red squares in Fig. S1b). Correspondingly, the zero magnetic field two-dimensional (2D) resistivity $\rho_{xx}$ value increases rapidly and reaches a peak in the charge neutral regime, which is displayed in blue in the Fig. S1b.

### B. Raw data of magnetoresistance under various gate voltages

In the main text, the relative change of the magnetoresistance (MR) is shown in the form of $\frac{\delta \rho}{\rho_0} = \frac{\rho_{xx}(H) - \rho_{xx}(0)}{\rho_{xx}(0)}$. The reason why we plot $\frac{\delta \rho}{\rho_0}$ instead of $\rho_{xx}$ itself is because $\frac{\delta \rho}{\rho_0}$ is a crucial parameter to the application in spintronics, such as MR sensors. Moreover, the absolute $\rho_{xx}$ value changes dramatically with varied gate voltages compared to the magnetic field dependence. So it is not very instructive to put all the $\rho_{xx}(H)$ curves in the same figure with the same scale. In this regard it is much clearer to display the relative change $\frac{\delta \rho}{\rho_0}$. To reveal the raw data of MR to the readers, in Fig. S2 of this session we display the magnetic field dependence of the absolute resistivity values measured at varied gate voltages at four representative temperatures.

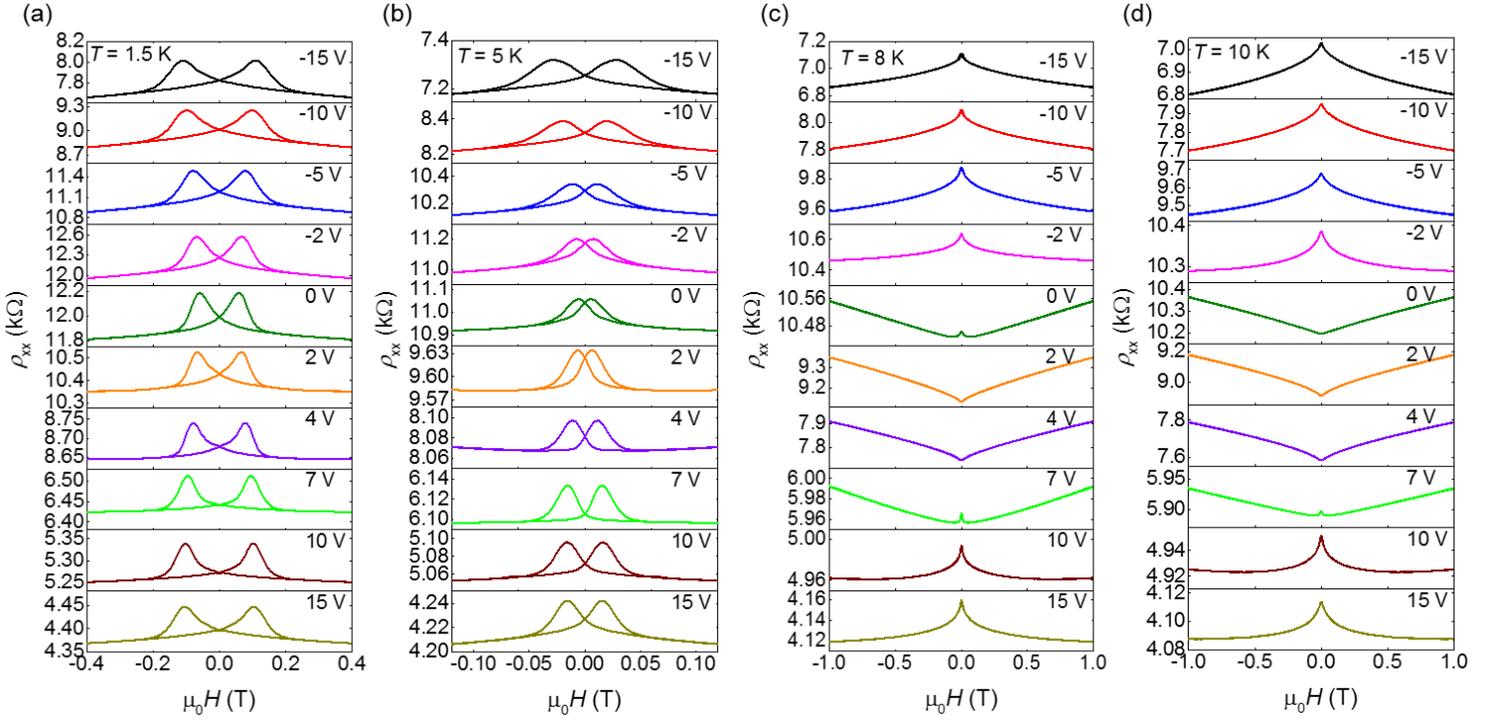

**Figure S2: The raw MR data measured under various gate voltages for *T* = 1.5 K to 10 K.** a,b, The MR taken at *T* = 1.5 K (a) and 5 K (b) show butterfly-shaped hysteresis, characteristic of a ferromagnetic material. The coercive field increases as more charge carriers are injected. c,d, Twice sign change of MR are clearly demonstrated at *T* = 8 K (c) and 10 K (d).

## C. Contour plot of the Anomalous Hall effect

The magnetization of a ferromagnetic thin film can be characterized by the anomalous Hall effect (AHE) because the total Hall resistivity is $\rho_{yx} = R_A M + R_H H$ [S2]. Here the zero field Hall resistivity $\rho_{yx}^0$ is proportional to the remnant magnetization of the sample. Fig. S3 shows the contour plot of the gate-tuned anomalous Hall effect based on the Hall traces shown in Fig. 1c of the main text. The $T_C$ can be defined as when $\rho_{yx}^0$ becomes zero with experimental uncertainties.

As indicated by the blue lines in Fig. S3, the general trend is also a "V"-shaped pattern. It is very similar to that obtained by the contour plot of $H_C$ shown in Fig. 2c of the main text based on the magnetoresistance measurements. As discussed in the main text, this trend implies enhanced ferromagnetic order on both *p*- and *n*-type regime when more carriers are injected by the electrical field.

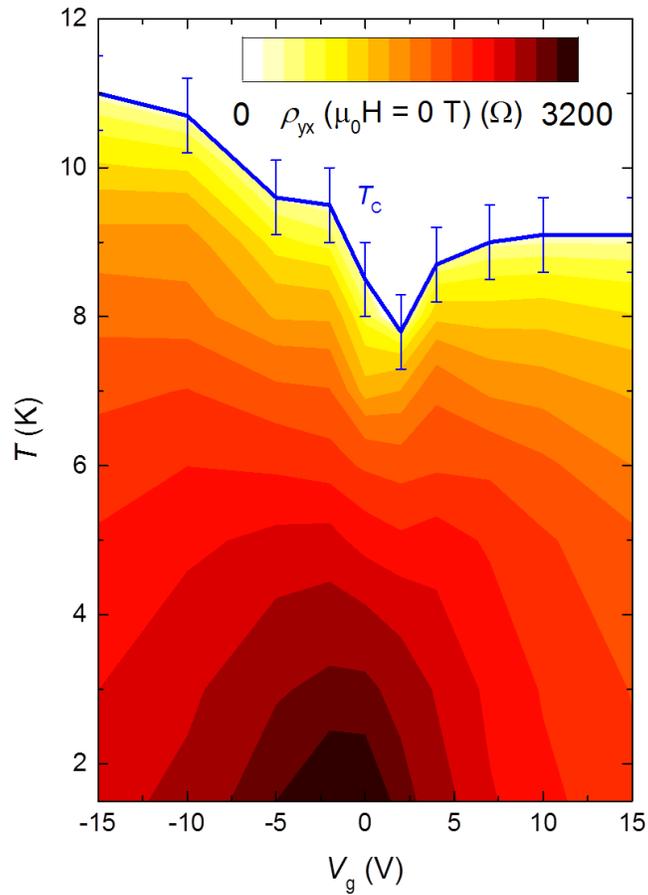

**Figure S3: The contour plot of the zero field Hall resistivity as a function of temperature and gate voltage.** The "V"-shaped $T_C$ line (blue) is also observed here, similar to that defined from the magnetoresistance measurements in Fig. 2c of the main text.

As temperature is reduced, the "V"-shaped pattern gradually evolves into a "dome" feature centered around -1 V. The reason why it is not centered around $V_g = +2$ V near the charge neutral point is not entirely clear. It might be related to the details of the magnetically opened gap below $T_C$, or the electronic band structure that determines the anomalous Hall conductivity. The strongly enhanced $\rho_{yx}^0$ (more than 3 k$\Omega$ at $T = 1.5$ K for $V_g = -2$ V to 0 V) is a manifestation that the system is not very far from the quantum anomalous Hall effect (QAHE) regime. However, the fabrication process involved here in making the FET device with top gate tends to degrade the quality of the film. Therefore the QAHE was observed in magnetically doped TI films without any capping layer. In that case only a bottom gate is applied across the STO substrate, which has a large dielectric constant at low $T$ [S3]. The advantage of the top-gate FET device used here is the ability to tune the carrier density over a large doping and temperature range, which allows us to observe the complex evolution of MR.

### D. Temperature dependent magnetoresistance measured at varied gate voltages

As discussed in the main text, the negative MR at $V_g = -15$ V (Fig. 4a) is caused by the magnetism-induced spin-dependent scattering of bulk carriers and the positive MR at $V_g = +2$ V (Fig. 4b) originates from the topology-induced WAL of the surface Dirac fermions. The MR behavior measured at the intermediate gate voltages can be understood as the competition between these two effects. Here in Fig. S4 we present the temperature-dependent MR curves measured at other gate voltages, which clearly demonstrate the smooth evolution between these two limits.

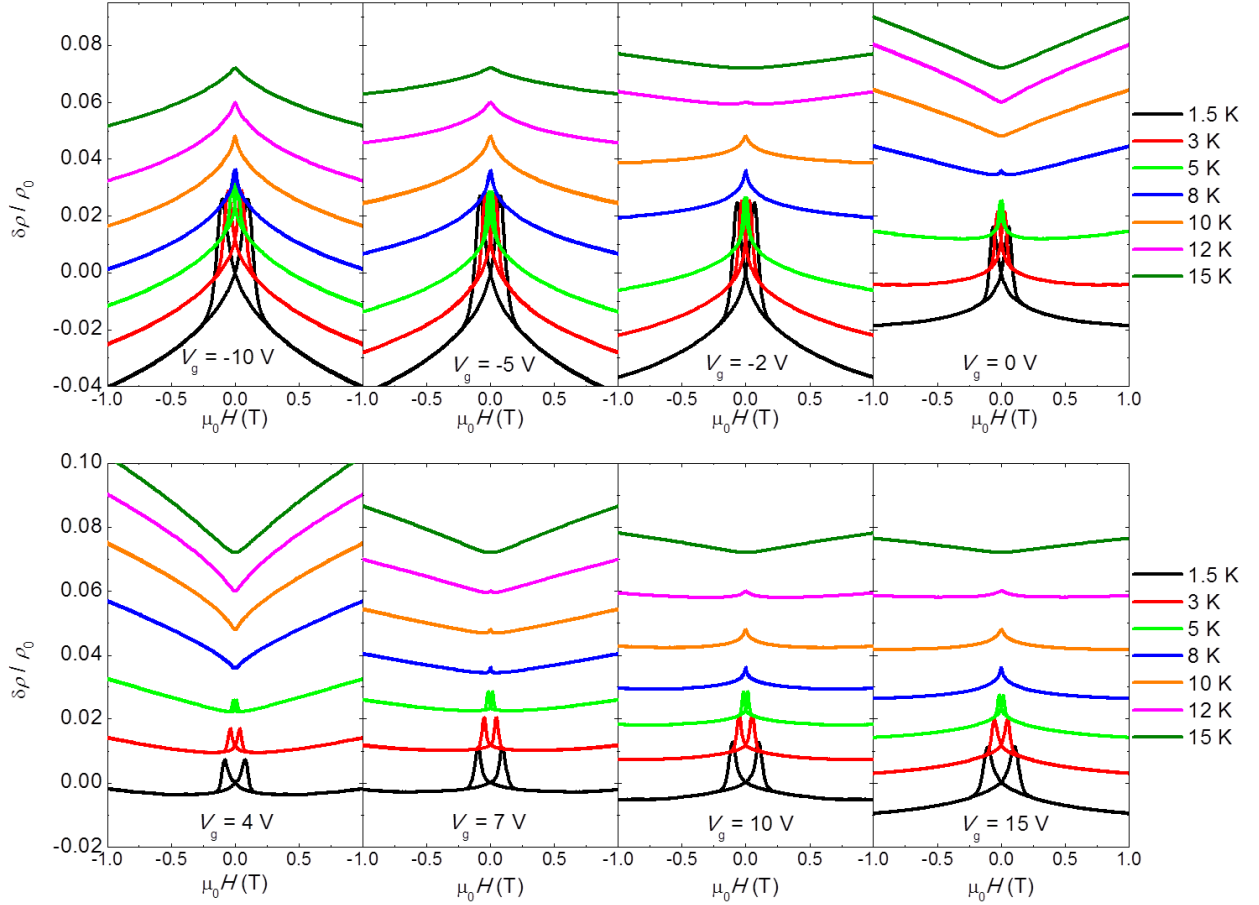

**Figure S4: Temperature dependent magnetoresistance measured at varied gate voltages.** With varied $V_g$ from -10 V to +15 V, the behavior evolves smoothly from the bulk-negative-MR-dominated regime to surface-WAL-dominated regime and back to the bulk-negative-MR-dominated regime via intermediate regimes (e.g. $V_g$ = -2 V and +10 V). For clarity each curve is offset by 1.2 percent.

For $V_g$ = -10 V and -5 V, the behavior is qualitatively the same as that measured at $V_g$ = -15 V because now the $E_F$ still lies below the Dirac point (DP) and bulk valence band holes dominate the magneto transport. Quantitatively the negative MR becomes weaker as $E_F$ moves towards the DP, because the ferromagnetic order is weakened with reducing hole density. The $V_g$ = -2 V belongs to the intermediate regime. The MR at $T < 10$ K is still strongly negative but above that it becomes positive. This is because now the bulk and surface hole-like carriers make comparable contributions to the transport. At $T$ below $T_C$, the strong magnetic order makes the negative MR more pronounced, and above $T_C$ the WAL of surface Dirac fermions becomes more important. The $V_g = 0$ V, +4 V and +7 V results are very similar to that measured at $V_g = +2$ V because in this regime the $E_F$ is very close to the DP so that the surface Dirac fermions dominate charge transport. When $V_g$ is increased to +10 V, the results are very similar to that taken at -2 V because both gate voltages correspond to the intermediate regime where the negative MR and WAL have comparable strength. For $V_g = +15$ V, the $E_F$ now lies in the bulk conduction band. The negative MR becomes dominant again, although it is not so strong as in the case of $V_g = -15$ V due to weaker FM order. Therefore the WAL of surface states is observation, especially at high temperatures (e.g. $T = 15$ K).

### E. Contour plot of the relative change of magnetoresistance

In Fig. 4e of the main text, we plot the temperature and gate voltage dependence of the relative change of MR measured at magnetic field at $\mu_0 H = 0.3$ T. The reason to choose this field is because it is larger than the coercive field at all temperatures and gate voltages so that the hysteresis effect can be avoided. Here in Fig. S5 we show two similar plots for the relative change of MR measured at $\mu_0 H = 0.5$ T and 1 T, respectively. The contour plots show the same "V"-shaped feature, where the negative MR appears in the regime of high $V_g$ and low temperature, which has stronger FM order. Therefore, the close correlation between the MR and FM order is a robust feature independent of the magnetic field, as long as it is larger than the coercive field of all temperatures and gate voltages.

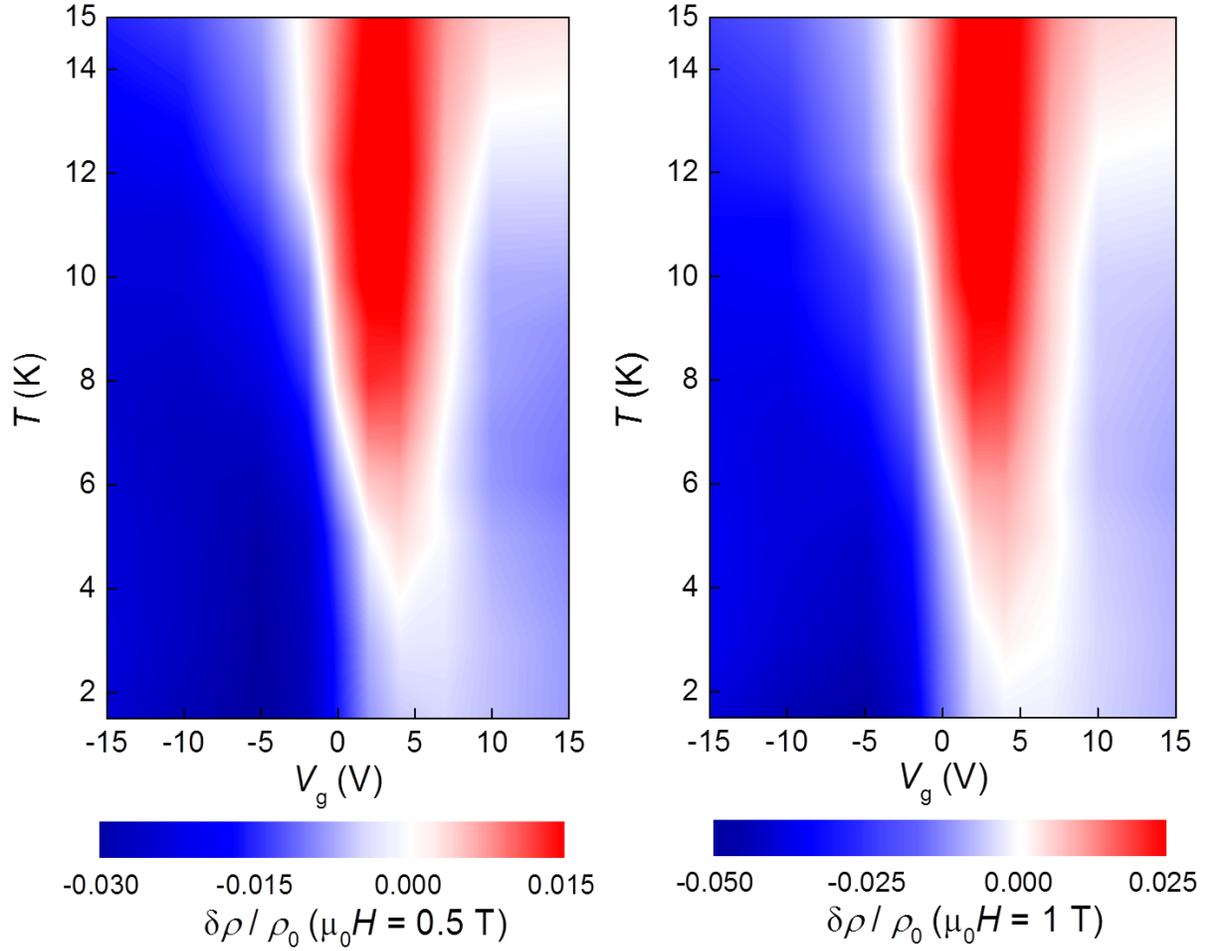

**Figure S5: Contour plot of the relative change of MR at magnetic field of 0.5 T (left) and 1 T (right).** The "V"-shaped feature is present in both plots, confirming that the close correlation between MR and FM order is a robust feature insensitive to the magnetic field.

## F. HLN formula fit for the magnetoconductivity curves

In the main text we propose that the negative MR in the large $|V_g|$ regime is due to the spin-dependent scattering of the bulk carriers rather than the weak localization (WL), which can also gives rise to a negative MR. This conclusion is drawn based on a thorough analysis of the detailed field dependence of resistivity. Below we provide both qualitative argument and quantitative analysis of the MR behaviour to justify this conclusion.

From a qualitative viewpoint, there is a smooth temperature evolution of the negative MR at $V_g$ = -15 V across $T_C$ = 11.5 K. As shown in Fig. 4a in the main figure, the butterfly-shaped negative MR is observed below 10 K, characteristic of the spin-dependent scattering in the FM state. At 15 K, just above $T_C$, the MR behavior is qualitatively the same except for the disappearance of the hysteresis. This is very reasonable because the ferromagnetic fluctuation is still very strong at such temperature as confirmed by the anomalous Hall effect (Fig. 1d in the main figure). Therefore, spin-dependent scattering should not disappear suddenly.

For a more quantitative analysis, we have used the well-known Hikami-Larkin-Nagaoka (HLN) formula [S4] for localization to fit the magnetoconductivity (MC) curves. By using the measured $\rho_{xx}$ and $\rho_{yx}$ curves, the MC can be obtained by $\sigma_{xx} = \frac{\rho_{xx}^2}{\rho_{xx}^2 + \rho_{yx}^2}$. For the TIs, the localization-induced MC can be fitted by the HLN formula with the assumption that $\tau_\phi \gg \tau_{so}, \tau_e$:

$$\delta\sigma(B) = \sigma(B) - \sigma(H) = \frac{\alpha e^2}{2\pi^2 \hbar} \left( \psi\left(\frac{1}{2} + \frac{\hbar}{4eBl_\phi^2}\right) - \ln\left(\frac{\hbar}{4eBl_\phi^2}\right) \right)$$

There are two fitting parameters in the HLN formula: the prefactor $\alpha$ reflects the type (WAL or WL) and amplitude of the corrections to conductivity, and $l_\phi$ is the phase coherence length of the charge carriers.

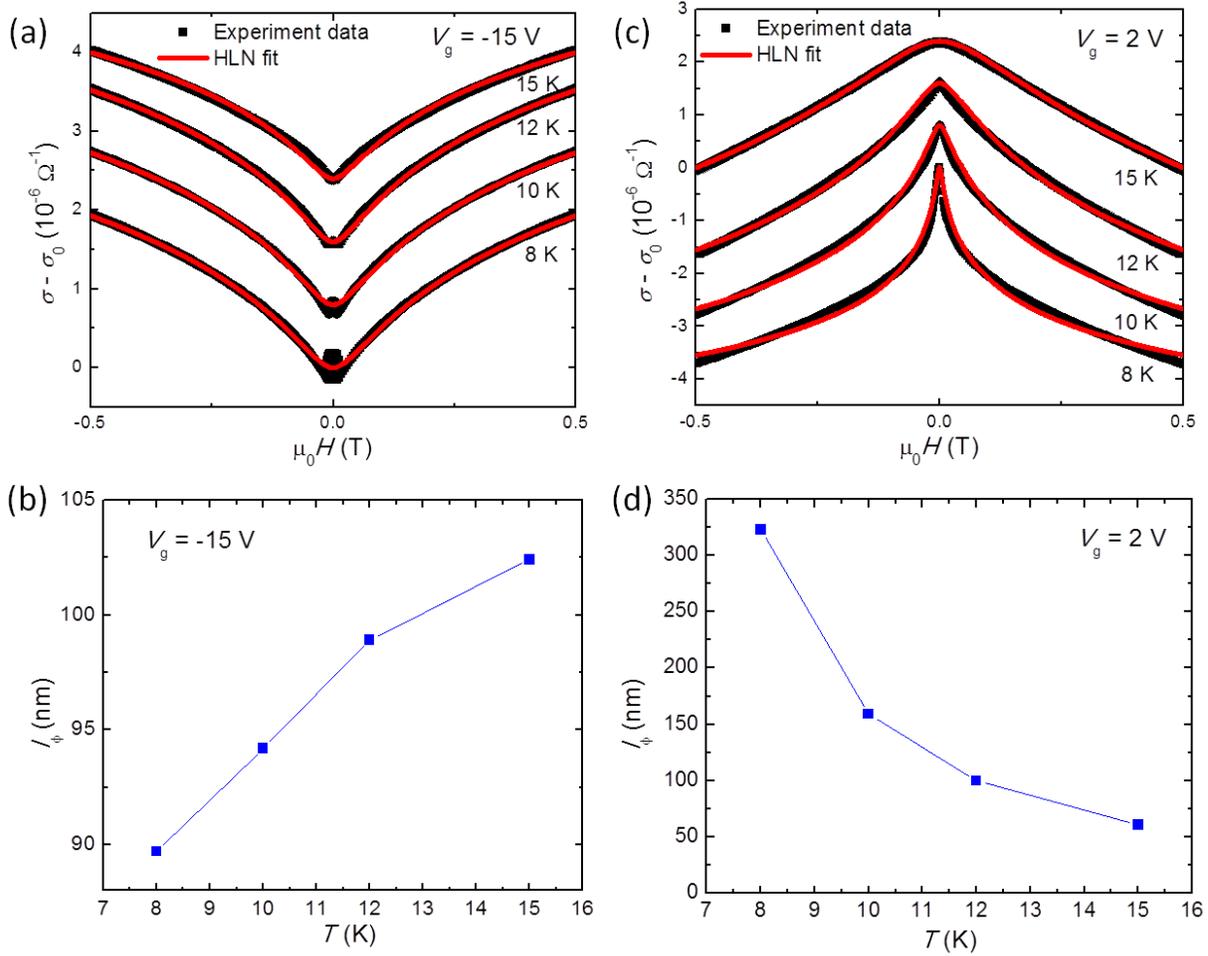

**Figure S6: HLN fit of the magnetocoductivity curves and the extracted coherence length.** (a) The shape of MC at $V_g$ = -15 V is insensitive to temperature. (b) Extracted coherence length increases with rising temperature, which is unphysical. (c) For $V_g$ = 2 V, the MC curves become broader at high temperature, indicating the decrease of coherence length. (d) The coherence length decreases with increasing temperature, consistent with the MC feature in Fig. S6c.

The fitting results for the curves measured at $V_g$ = -15 V, in which bulk holes dominate the transport, are plotted in the Fig. S6a. Although the curves can be fitted fairly well by the WL formula, a closer examination of the data reveals that the temperature evolution of the fitting parameters cannot be right. As shown in Fig. S6b, the extracted phase coherence phase $l_\phi$ at $V_g$ =

-15 V increases systematically from 89.7 nm at 8 K to 102.4 nm at 15 K, which is unphysical because $l_\phi$ is determined by the inelastic scattering, which should decrease at higher temperatures due to enhanced inelastic scattering rate. This trend can actually be directly visualized from the data without fitting. From 8 K to 15 K, the shape of the MC curve at $V_g$ = -15 V is almost identical, indicating that the characteristic field scale (hence the length scale) is not changing significantly with temperature. The reason why the $l_\phi$ in the HLN fit increases at higher temperature is because the prefactor $\alpha$ decreases at higher temperature due to the reduced amplitude of MC. This is strong evidence to against the WL mechanism for the negative MR, but is consistent with the spin-dependent scattering due to ferromagnetic fluctuations.

In contrast, the MC curves taken at $V_g$ = 2 V can be fitted very well with reasonable parameters by using the HLN formula for WAL. As shown in Fig. S6c, the MC curves become broader at higher temperature, indicating the increase of field scale and thus decrease of length scale. Accordingly, the extracted $l_\phi$ value decreases with increasing temperature due to the enhanced inelastic scattering rate. It should be noted that the fitting curve is becoming less satisfactory as the temperature is approaching Currie temperature $T_c$ from high temperature. This again confirms the enhanced magnetic fluctuation at low temperature.